\documentclass{article}
\usepackage{amsfonts}
\newcommand{\beq}{\begin{equation}}
\newcommand{\eeq}{\end{equation}}
\begin{document}

\title{\bf Many--body interactions among adsorbed atoms and molecules
within carbon nanotubes and in free space}

\author{Milen K. Kostov\footnote{\em corresponding author, e-mail: mkk143@psu.edu, fax: 814-865-3604}\, ,
  Milton W. Cole, and
  John Courtenay Lewis \\
{\em Department of Physics,} \\
{\em The Pennsylvania State University, University Park, PA 16802, USA}
\and
  Phong Diep and J. Karl Johnson \\
{\em Department of Chemical and Petroleum Engineering,} \\
 {\em The University of Pittsburgh,
 Pittsburgh, PA 15261, USA}}

\maketitle

\begin{abstract}
Studies of three--dimensional
            and two--dimensional condensed phases have shown that many--body interactions       
 contribute $\sim10\%$
 to the equations of state of noble
 gases. This paper assesses the importance of three--body triple dipole interactions for quasi--one--dimensional phases of He, Ne, H$_2$, Ar, Kr and Xe  confined within
 interstitial channels or on the external surfaces of nanotube bundles.
 We find the
 substrate--mediated contribution to be substantial:
for interstitial H$_2$ the well depth of the effective pair
 potential is reduced to approximately one half of its value in free space.
 
 We carry out {\em ab initio} calculations on linear and equilateral triangular configurations of (H$_2$)$_3$
 and find that overlap interactions do not greatly change the DDD interaction in the linear configuration when the lattice spacing is greater than about $3$ \AA. However, the DDD interaction alone is clearly insufficient for the triangular configurations studied.

\end{abstract}

\section{Introduction}

Adsorbates within carbon nanotubes and bundles of nanotubes have been the focus of
much recent attention (see for example refs. \cite{sim}--\cite{cole}). In particular, novel quasi--one
dimensional adsorbate phases have been predicted. The closest to true one--dimensional
behavior is likely to be realized by small atoms or molecules (He, Ne, H$_2$) in
the ``interstitial channel'' (IC) phase, and the ``groove'' phase. For the IC phase,
the molecules adsorb in channels bounded by three contiguous nanotubes within bundles.
The groove phase comprises adsorbate in the grooves between adjacent tubes on the
perimeter of a nanotube bundle. In both of these phases the molecules are confined
transversely but can move more or less freely along the lengths of the nanotubes.

To date, in exploring adsorption on and in nanotubes, relatively little attention
has been paid to the detailed nature of the interactions among adsorbate molecules.
Indeed, most studies have assumed that the pairwise interaction does not differ
significantly from its free--space form. In this paper we show that, for the IC and
groove systems considered, the proximity of a polarizable medium, the nanotube walls,
will profoundly alter the effective pair potential between two adsorbate molecules,
to the extent of producing qualitative changes in expected behavior.

In two-- and three--dimensional systems many--body effects have been extensively studied~\cite{buck,bark,book}. Barker~\cite{bark}, in particular, showed that equilibrium properties of condensed rare gas systems could be obtained accurately using potentials in which the pair potentials were obtained from gas--phase data, and the higher--order contributions were purely of Axilrod--Teller--Muto or ``triple--dipole'' (DDD) type, even though this is exact only asymptotically for large molecular separations. Barker argued that extensive cancellation must take place among other many--body interactions, at least one of which is known to be comparable in magnitude to the DDD interaction.

Analogously in two dimensions, adsorbed monolayer films manifest two types of
many--body contributions that are relevant to quasi--one--dimensional nanotube phases.
One is that involving three adatoms. The other is a modification of the adatom pair
potential by the substrate acting as a third body. McLachlan derived an asymptotic
expression, valid at large adatom separation, for this substrate mediation energy~\cite{mcl}.
As in the three--dimensional case, experimental data at condensed phase densities in
adsorbate films are consistent with the asymptotic expression~\cite{raub,gott}. However, the
experimental data for two--dimensional films are too limited to provide a thorough
test of the adequacy of McLachlan's interaction. We note that a third three--body
energy in films involves two substrate atoms and an adatom. This oft--neglected
interaction can make a substantial correction to the gas--surface adsorption potential.

While extensive cancellation occurs as discussed above in three dimensions, among
many--body interactions other than the DDD three--body interaction, this cannot be assumed
{\em a priori} in one dimension because of the strong angular dependence typical of many--body
interactions. We have therefore extended a recent accurate {\em ab initio} calculation of the
potential energy surface of H$_2$--H$_2$ \cite{Diep2} to linear equispaced (H$_2$)$_3$ and to (H$_2$)$_3$ arranged
in equilateral triangles. While significant deviations from the DDD interaction are
found for the equilateral
triangular configurations, the deviations from DDD for linear (H$_2$)$_3$ are important only for
separations small compared to the equilibrium separation of H$_2$--H$_2$.

In this paper we consider a linear array of rare gas atoms or of H$_2$, henceforth denoted A,
each separated from its nearest neighbor by a lattice constant $a$. In Section $2$ we
evaluate the net DDD contribution arising from interactions among all sets
of triplets AAA. The resulting three--body energy per atom, which we call $E^{(3)}_{AAA}$,
turns out to be negative and of order $1\%$ of the total pair interaction energy.
We next evaluate an energy per pair $V^{(3)}_{AAC}$, due to a summation of the DDD interactions
between two A particles and all C atoms. This energy is found to be positive and of
large magnitude, equal approximately to half of the interaction between the atoms
in free space. When $V^{(3)}_{AAC}$ is added to the free space interaction, there results
an effective pair potential which has a well depth significantly smaller than
the free space well depth. This reduced attraction implies that collective
phenomena\footnote{In this paper we do not consider the implications of substrate--mediated
adsorbate interactions for
condensation involving interactions among {\em different} IC or groove phases in a nanotube bundle (see e.g. Ref.\cite{cole}).},
such as condensation, ought to occur at much lower temperatures than would be predicted
if only the free--space interactions were taken into account, and substrate mediation
were ignored. Qualitatively similar behavior is derived as a consequense of the DDD interaction for  He, Ne, H$_2$, Ar, Kr and Xe atoms/molecules adsorbed into the grooves of a nanotube bundle. In Section $3$ we discuss the {\em ab initio} calculation
carried out on (H$_2$ )$_3$ in linear and triangular arrays.

While this paper evaluates electronic substrate mediation of adsorbate interactions,
a substrate phonon mediation mechanism also exists. Such phonon mediation effects on adatom interactions on planar surfaces are reviewed by Gottlieb and Bruch \cite{gott}. While no such evaluation has been performed for atoms or
molecules in nanotubes, the analogous effect of nanotube phonons on the
electron--electron interaction has recently been assessed by Woods and Mahan~\cite{woods}.

\section{Adsorbates in IC and Groove Phases}

We begin by considering particles A confined to the axis of an IC phase and separated by a
lattice constant $a$. There are two relevant three--body energies which we consider. The
first is the net DDD energy arising from interactions among all
sets of triplets on the line. The second is the total DDD interaction involving an adatom pair
and all possible C atoms. The potential energy $\Phi$ corresponding to a
linear array of $N$ particles specified by their position vectors $\vec{r_1},...,\vec{r_N}$
is given by 
\beq
\Phi(\vec{r_1},\ldots,\vec{r_N}) = \sum_{i<j}^{N}U^{(2)}_{AA}(\vec{r_i},\vec{r_j}) +
\sum_{i<j<k}^{N}U^{(3)}_{AAA}(\vec{r_i},\vec{r_j},\vec{r_k}) + \sum_{i<j<k}^{N}U^{(3)}_{AAC}(\vec{r_i},\vec{r_j},\vec{r_k})\, ,
\eeq
where $U^{(2)}$ and  $U^{(3)}$ are pair and DDD triplet potential
functions and four-body (and higher order) interactions are ignored. For purposes of estimating the total pair energy we will temporarily
assume that the pair potential $U^{(2)}_{AA}(\vec{r_i},\vec{r_j})$ is isotropic and of
Lennard--Jones 12--6 form with distance parameter $\sigma$
and energy parameter $\epsilon$.
The total pair interaction energy per particle A can be written as
\beq
E^{(2)}_{AA} = \frac{V^{(2)}_{AA}}{N} = 4\epsilon\left[ \left( \frac{\sigma}{a}\right)^{12}\zeta(12)
- \left( \frac{\sigma}{a} \right)^6\zeta(6)\right]\, ,
\label{eq,pp}
\eeq
in the limit $N \rightarrow \infty$, where $\zeta$ is the Riemann zeta function.
The right--hand side of Eq. (\ref{eq,pp}) has a minimum at $a/\sigma = 1.122$.
Evaluating Eq. (\ref{eq,pp}) at that minimum we obtain:
\beq
E^{(2)}_{AA} = \frac{V^{(2)}_{AA}}{N} = -1.035 \epsilon =  - 0.509\frac{C_6}{a^6}\, ,
\eeq
where $C_6=4\epsilon\sigma^6$.
Next we assess the contribution of the DDD interactions $U^{(3)}_{AAA}$ to $V^{(2)}_{AA}$. A simple result
for $U^{(3)}_{AAA}$ holds for isotropic oscillators, at large separations:
\beq
 U^{(3)}_{AAA} = \nu_{AAA}\frac{3 \cos(\theta_i)\cos(\theta_j)\cos(\theta_k) + 1}
{r^3_{ij}r^3_{ik}r^3_{jk}}\, ,
\label{eq:aaa}
\eeq
where  $\nu_{AAA}$ is the triple dipole dispersion energy coefficient, $ r_{ij}, r_{ik}, r_{jk}$ are the interatomic
distances in a given triplet, while $ \theta_i, \theta_j$ and $\theta_k$ are the internal angles of the triangle
formed by the atoms $i, j$ and $k$. The sign of the triple dipole energy
depends on the geometry of the triangle formed by the three atoms. Indeed, the DDD term is positive for acute
triangles, while for most obtuse triangles it is negative. In particular, $U^{(3)}_{AAA}$ is negative for the linear configuration of adsorbate particles, since $\cos(\theta_i)\cos(\theta_j)\cos(\theta_k) \simeq -1 $ in that case.
Then, the net DDD potential becomes:
\beq
V^{(3)}_{AAA}\cong - \sum_{i<j<k}^{N} \frac{2  \nu_{AAA}}{( r_{ij}r_{ik}r_{jk} )^3}\, .
\label{eq,thbaaa}
\eeq
Summing over all possible triplets  we obtain the net DDD energy per atom:
\beq
E^{(3)}_{AAA} = -\frac{2  \nu_{AAA}}{a^9} \left[ \frac{1}{2^3}\zeta(9) +
2 \sum_{n=2}^{\infty}\sum_{m=1}^{\infty} \frac{1}{m^3(n+m-1)^3(n+2m-1)^3} \right] \, ,
\label{eq,aaa1}
\eeq
\beq
E^{(3)}_{AAA} = - \frac{0.27 \nu_{AAA}}{a^9}\, .
\eeq
The coefficient $ \nu_{AAA}$ can be estimated from a Drude  model of isotropic oscillators from the coefficient $C_6$ : $ \nu_{AAA} = \frac{3}{4}C_6\alpha_{A}$, 
where $\alpha_{A}$ is the static polarizability of the adatom. Then the ratio of the DDD
energy to the  pair interaction energy is:
\beq
\frac{E^{(3)}_{AAA}}{E^{(2)}_{AAA}} \cong {0.4\frac{\alpha}{a^3}}\, .
\eeq
This ratio is very small ($0.8\%, 0.3\%$ and $0.5\%$ for H$_2$, He and Ne, respectively, using values of $\alpha$ and {\em a} taken from \cite{book}).


We next 
evaluate and assess the net DDD interaction involving the adatom pair and all possible C atoms.
The effective pair potential has the form:
\beq
V_{\mathrm{eff}}^{(2)}(\vec{r_i},\vec{r_j}) =  U^{(2)}_{AA}(\vec{r_i},\vec{r_j}) +  
V^{(3)}_{AAC}(\vec{r_i},\vec{r_j})\, .
\eeq
For a linear configuration of adsorbate particles:  $ U^{(2)}_{AA}(\vec{r_i},\vec{r_j}) = U^{(2)}_{AA}(|z_i-z_j|)$ and $V_{\mathrm{eff}}^{(2)}(\vec{r_i},\vec{r_j})=\\ =V_{\mathrm{eff}}^{(2)}(|z_i-z_j|)$, where $z$ axis is along the axis of the IC. The DDD energy $V^{(3)}_{AAC}$ is given by:
\beq
V^{(3)}_{AAC}(\vec{r_i},\vec{r_j}) = \sum_{k}U^{(3)}_{AAC}(\vec{r_i},\vec{r_j},\vec{r_k})\, .
\label{eq,atm}
\eeq
Here $\vec{r_k}$ is the position of the $k^{th}$ carbon atom along
the nanotube surface and $U^{(3)}_{AAC}$ is its DDD interaction with the
adatoms at $\vec{r_i}$ and $\vec{r_j}$. In the approximation of isotropic
oscillators, the  expression  for the DDD energy is analogous to Eq.(\ref{eq:aaa}), except that the strength coefficient is:
\beq
\nu_{AAC} = \frac{3\alpha^2_{A}\alpha_{C}E_{C}E_{A}(E_{C}+2E_{A})}{4(E_{A}+E_{C})^2}\, .
\label{eq,nu}
\eeq 
Here the energies $E_{A}$ and $E_{C}$ are characteristic energies of the adatom and C atom,
respectively, and $\alpha_{C}$ is the static polarizability of a C
atom \cite{raub,klein}. 

For simplicity, we perform the summation of gas-surface DDD interactions (\ref{eq,atm}) by smearing out
the carbon atoms in the substrate. This continuous
approximation introduces an error, but the qualitative trends are expected to be accurate in general. Under this assumption,
the net DDD contribution from three nanotubes of radius $R$ takes the form:
\beq
 V^{(3)}_{AAC}(z) = \frac{12\vartheta\nu_{AAC}M(x ; y)}{z^3R^4}\, ,
\label{eq,1cn}
\eeq
where $\vartheta = 0.38$ \AA$^{-2}$ is the surface density of C atoms
and $z$ is the distance between adatoms. Here, $\displaystyle{x=z/R}$ and $\displaystyle{y=d/R}$, where $d$ is
the distance from the axis of the cylinder to the axis of the 
IC. The integral $M(x ; y)$ is a dimensionless integral over a cylindrical surface. Results for $V^{(3)}_{AAC}$ as a function of adatom lateral separation, in the case of
quasi-$1$d phases of He, Ne and H$_2$ adsorbates, were calculated from (\ref{eq,1cn}) for nanotubes of radius $6.9$ \AA\, 
and $d=9.815$ \AA.

The pair potentials employed here are those of Silvera and Goldman \cite{silv} (H$_2$); Aziz, Nain {\em et al.} \cite{aziz} (He) and of Aziz, Meath {\em et al.} \cite{aziz1} (Ne). The values of DDD strength coefficient (\ref{eq,nu}) were calculated from Ref.\cite{raub,klein}. Results of our calculations are given in Fig. $1$, Fig. $2$ and Fig. $3$  
and in Table $1$.  In Table $1$, $\epsilon$ and $z_m$ are the well depth and the location of the
potential minimum for  $V_{\mathrm{eff}}^{(2)}$; $\epsilon_0$ and $z_{0m}$ are the corresponding quantities
for the free space potential; $\delta$ is the percent decrease in the well depth due to $V^{(3)}_{AAC}$  and $\rho$
is the sign-reversal point of  $V_{\mathrm{eff}}^{(2)}$.

In all three cases we observe a  large repulsive DDD effect: the reduction in the well depth of the pair interaction
is $54\%$, $28\%$ and $25\%$ for H$_2$, He and Ne, respectively (Table $1$). An interesting feature of the effective
potential is the presence of a critical value of adatom separation for H$_2$ and He adsorbates, beyond which  $V_{\mathrm{eff}}^{(2)}$ becomes repulsive for a certain range of separation values. These points are
$\rho\simeq 6.3$ \AA\  and  $7.4$ \AA\  for H$_2$ and He, respectively (Table $1$), with corresponding ranges of repulsive  $V_{\mathrm{eff}}^{(2)}$: $\triangle z$=($6.3 - 15.1$ \AA ) and $\triangle z$=($7.4 - 11.7$ \AA ). On the other hand, an investigation of the asymptotic behavior of $M(x ; y)$ reveals that the DDD energy becomes negative for very large interparticle separations ( $z > 9R $ ). We note that a reversal in sign for  $V_{\mathrm{eff}}^{(2)}$ is not found for Ne. To elucidate this fact we investigate the relative strength
of DDD interaction at $z_{0m}$. The expression (\ref{eq,1cn}) can be used to yield:
\beq
\frac{V^{(3)}_{AAC}(z_{0m})}{\epsilon_0}\equiv kM(x_{om} ; y)[\frac{\nu_{AAC}z_{0m}^3}{C_6}]\, ,
\label{eq,inv}
\eeq
where $\displaystyle{\epsilon_0\simeq C_6/2z^6}$ is the free space well depth and $k$ is independent of the adsorbate species. The integral
$M(x_{0m} ; y)$ has only a weak dependence on $z_{0m}$ in the region of interest ($3-3.5$ \AA).
Then the relatively small DDD effect for Ne is explained by the bracketed term involving $\nu_{AAC}$, 
$C_6$ and $z_{0m}$ in (\ref{eq,inv}). Its value is nearly twice as large for H$_2$ as for Ne ( $34.1$ \AA$^6$ \, vs.\,  17.7 \AA$^6$ ).

The present results imply that the substrate mediation effect is
approximately three times as large for the IC geometry than for
adsorbed monolayer films \cite{book}. Two factors are responsible.
One is that the equilibrium distance to the substrate ($2.9$ \AA)  is $\sim 5\%$ smaller than on a planar surface, yielding a larger substrate contribution.
The more important factor is that three adjacent surfaces contribute in
this IC geometry.

One significant consequence of the reduced attraction between the particles
within an IC is a qualitative change in the predicted equation of state for
this 1d system. It has been shown \cite{krot} that the
ground state of a 1d bose system is a liquid if and only if a bound dimer
exists. For He, the dimer is bound by just a few milliKelvin, if the free
space potential is used \cite{stan3,bonin,gord}. There is no doubt therefore that this
binding disappears when the weakened attraction is substituted. For H$_2$, the
dimer binding energy computed with the Silvera-Goldman potential is large (0.384 meV) and the ground state of the system is a liquid with cohesive energy $\simeq 0.4$ meV per molecule \cite{gord1}. When the new potential appropriate to the IC is used, we find no
bound states. ( For D$_2$, the dimer is very weakly bound, by 0.0513  meV ). If our
calculations are correct, they imply that the ground states of interstitial $^4$He and
para-H$_2$ are quantum gases. Such a system is of considerable fundamental
interest. 

%
%
    
In the nanotube bundle geometry, adsorption can also occur on the outer
surface of a bundle. The most favorable site energetically is the groove
formed at the intersection of two external tubes, which has been studied by
several groups both experimentally and theoretically. In contrast to the IC case, larger particles ( such as Ar, Kr and Xe ) can fit nicely into the grooves \cite{stan}.
 We have computed the
three-body effect of the substrate, using the DDD expression as in the
IC's. The resulting effect on the pair potential for H$_2$, He, Ne, Ar, Kr and Xe is reported in Table $2$.
The substrate effect of reducing the well depth is smaller than in the	
IC case. The reasons are twofold. One is simply that only two nanotubes
contribute here, instead of three in the IC case. The second reason is that
the distance to the tubes is somewhat larger than in the IC case; in these
grooves, the distance is very close to the $\displaystyle{\sigma}$ parameter in the
gas-carbon pair interaction. (It would be exactly $\displaystyle{\sigma}$ if the tube radius
were infinite). These differences parallel those used above to distinguish
the IC case from the graphite case. For this reason, the groove results are
intermediate between those from the latter two geometries.



\section{Three-body interactions for the H$_2$ trimer in free space}
We have calculated the nonadditive, three-body
interactions for the linear and equilateral triangle arrangements
of three hydrogen molecules from first principles.
The H$_2$ bond length is kept fixed at 
0.7668 \AA, consistent with the rigid rotor
approximation.
With the molecular centers-of-mass forming the vertices of a triangle,
the intramolecular orientation of each molecule is described relative
to three axes centered at each vertex.  
We have explored the effects of molecular orientation on 
the three-body interactions by orienting each molecule in one of 
three ways. Each molecule is oriented with its molecular axis perpendicular
to the plane of the triangle, bisecting a corner of the triangle,
or perpendicular to the line bisecting a corner.
In total, there are 27 possible combinations of trimer ``orientations''.
Due to symmetry, 
it is necessary to calculate only 10 non-degenerate orientations
for each geometry.
The closest center-of-mass to 
center-of-mass distance was stepped by 0.2 angstrom from
2.0 to 6.0 angstrom for the linear geometry and from 3.0
to 6.0 angstrom for the equilateral triangle arrangement.

Convergence tests were performed to determine the level of theory and basis
set required for accurate calculations of three-body hydrogen interactions.
We found that 
the coupled-clusters 
(CCSD(T)\cite{Gauss}) 
level of theory with Dunning's augmented, correlation consistent, 
triple zeta basis set, aug-cc-pVTZ\cite{Dunning}
was sufficient to converge the three-body energies to within 0.1 K.
The generalized counterpoise method of Valiron and Mayer\cite{Valiron2}
was used to correct for basis set superposition error (BSSE).

In order to compare with the three-body contributions calculated using
the triple-dipole approximation for isotropic oscillators (Eq.4), 
we have calculated isotropic potentials for both the linear and 
equilateral geometries.  The isotropic potential is obtained by
weighted averaging of the potentials representing each of the 
ten combinations of orientations.  The weight placed on each
potential is proportional to the degeneracy of the specific orientation. 
In effect, this corresponds to uniformly averaging over all 27 possible
orientation.  This is illustrated for the equilateral
geometry in Fig. $4$.  

At close range,
Hartree-Fock level contributions such as 
electron exchange and repulsion effects can be 
large \cite{Rosen,Lotrich2,Higgins}.  Only at larger intermolecular
separations do dispersion forces dominate.  With this in mind,
we calculated the Hartree-Fock nonadditive three-body
potential energy surface 
using the same basis set and BSSE correction.  A 
``dispersion only'' three-body potential surface was approximated
by subtracting the Hartree-Fock PES from the original CCSD(T) surface.
Fig. $5$ shows the resulting dispersion potentials for the 
equilateral (filled symbols) and linear (empty symbols) geometries.    
This approach is not a true calculation of the dispersion potential.
But, the quickly decaying HF potentials
suggest that the remaining interactions are mainly dispersion.  
Fitting the
long range energies ($r_{\mathrm{nearest}}$ $\geq$ 3.8 \AA\ for the linear and
$r_{\mathrm{nearest}}$ $\geq$ 4.8 \AA\ for the equilateral) using the leading
triple-dipole dispersion function given by Eq. (4) 
resulted 
in a root-mean-square error of 0.013.  By including the next leading
term, 
\begin{equation}
\label{eqn:c3_wddq}
U(DDQ)=Z(DDQ)\frac{9\cos(\phi_3)-25\cos(3\phi_3) +
6\cos(\phi_1-\phi_2)\left[3+5\cos(2\phi_3)\right]}
{r_{12}^{3}r_{23}^{4}r_{31}^{4}},
\end{equation}
one further reduces the error to 0.006.
We did not find significant improvements to the fit by including the
remaining triple dispersion terms.  The dispersion model fitted to
both the linear and equilateral geometries are plotted in Fig. $5$
as dashed and solid curves, respectively. 
The resulting dispersion energy constants are 
$Z(DDD) = 49607$ K$\cdot$\AA $^9$ and 
$Z(DDQ) = 5763$ K$\cdot$\AA $^{11}$. 
McDowell and Meath report an ``exact'' value for $Z(DDD)$
of 49833 K$\cdot$\AA $^9$, 
calculated from dipole oscillator strength distribution 
data \cite{Meath1}.    
McDowell\cite{mcd} observed that for a several specific
orientations of H$_2$ trimers (all three molecules oriented in
the same plane), nonadditive induction contributions can be
as large as those from DDD.  This, however, is not apparent
in the isotropic potentials, where the induction effects 
have been completely ``washed-out'' by the
uniform angle-averaging process \cite{Lotrich2}, as is the case with permanent
multipole interactions.  

The Hartree-Fock potentials in Fig. $5$ 
indicate that non-dispersion interactions, such as exchange/repul-
sion effects,
are substantial at short range.  We will present a thorough analysis
of the different origins of hydrogen three-body interactions in a forthcoming 
article \cite{Diep3}.  It is clear from the present study that the
DDD approximation, alone, is not sufficient at describing the total (CCSD(T))
three-body interactions at short distances.  Whether
this hold true for systems containing the other constituents remains unclear
and would require separate {\it ab initio} calculations.

\section{Discussion}

In this paper we have found that three-body interactions play an important role in 
altering the interactions between adatoms within IC's between carbon nanotubes. The reduction
in the well depth of the pair interaction is $54\%$, $28\%$ and $25\%$ for H$_2$, He and Ne,
respectively. The magnitude of these shifts is roughly similar to the result of a crude estimate,
 i.e. a factor of three increase over the effect of a single planar graphite surface.
The DDD term even becomes larger in magnitude than the two-body potential for
interparticle separations above $\sim$  $7.4$ \AA\ and $6.3$ \AA\ for He and H$_2$, respectively.
Such a large DDD effect implies a significant change in the condensation properties of H$_2$
and He in carbon nanotube bundles. For (H$_2)_2$ the reduction in well depth is sufficient as to destroy the bound state. The perimeter of a nanotube bundle is accessible for adsorbate particles of bigger size.
In our approach, the observed reduction in the well depth of the pair interaction due
to DDD effect is still significant:
$35\%$, $24\%$, $15\%$, $19\%$, $22\%$ and $28\%$ for H$_2$, He, Ne, Ar, Kr and Xe, respectively.

The important consequence from the DDD effect for quasi-$1$d IC phase of H$_2$ molecules and He atoms
is that there is no  dimer bound state. The lack of the latter implies that there is no
bound state for the $N$ particle system. Then, the ground state of H$_2$ molecules
interacting with the screened pair potential is a gas, not a liquid.
This result raises an interesting question: what is the low temperature behavior of such a marginally unbound system?

Finally, {\em ab initio} calculations  of three-body interactions for linear and equilateral triangle arrangements of three H$_2$ molecules provide implications for the behavior of three-dimensional phases of hydrogen. Moreover, they justify a future {\em ab initio} study of the exchange interaction of two H$_2$ molecules and a carbon atom within a nanotube.

We are grateful to Keith Williams, Paul Sokol, Peter Eklund and Moses Chan for helpful discussions. This research has been supported by the Army Research Office under grant DAAD $19-99-0167$.

\newpage
\begin{center}
\Large{\bf Tables}
\end{center}
\begin{center}
\begin{table}[h]

\caption{DDD effect for H$_2$, He and Ne adsorbates confined within IC's of a nanotube bundle. All lengths are in \AA\, units;\, $\epsilon$ and $\epsilon_0$ are in meV.}
\setlength{\tabcolsep}{1cm}
\renewcommand{\arraystretch}{2}

\begin{tabular*}{13.96cm}[b]{|c|c|c|c|} \hline
Quantity&                      H$_2$&               He&                Ne\\ \hline
$\epsilon( \epsilon_0 )$&   1.35(2.96)&          0.67(0.93)&       2.72(3.64)\\
$\delta(\%)$&                54&                   28&               25\\
$z_m( z_{0m} )$&            3.55(3.41)&          3.00(2.97)&    3.12(3.09)\\
$\rho$&                       6.3&                 7.4&               -\\ \hline
\end{tabular*}

\end{table}  
\end{center}
\newpage
\begin{center}   
\begin{table}[h]
\caption{DDD effect for H$_2$, He, Ne, Ar, Kr and Xe adsorbates confined into the groove of a nanotube bundle. All lengths are in \AA\, units;\,  $\epsilon$ and $\epsilon_0$ are in meV. The free space potential parameters for Ar, Kr and Xe were taken from Ref. \cite{kle}.}

\setlength{\tabcolsep}{0.4cm}
\renewcommand{\arraystretch}{2.5}
\begin{tabular*}{16.125cm}[t]{|c|c|c|c|c|c|c|}\hline
Quantity&    H$_2$&    He&    Ne&    Ar&     Kr&    Xe\\ \hline

$\sigma$&              2.97&   2.74&  3.00&  3.40&   3.42&  3.40\\
$\epsilon( \epsilon_0 )$& 1.93(2.96)& 0.71(0.93)& 3.08(3.64)& 10.1(12.3)& 13.5(17.2)& 17.5(24.2)\\
$\delta(\%)$&         35&     24&    15&    19&     22&    28\\
$z_m(z_{om})$&   3.49(3.41)&   3.00(2.97)&  3.11(3.09)&  3.79(3.76)&   4.05(4.01)&  4.42(4.36)\\ \hline

\end{tabular*}

\end{table}

\end{center}

\newpage
\begin{center}{\bf Figure Captions}
\end{center}

Fig.$1$. Free-space potential $U^2_{AA}$ (circle/full  curve), the DDD
potential $V^{(3)}_{AAC}$ (triangle/full curve) and the effective pair
potential $V_{\mathrm{eff}}^{(2)}$(solid curve) for H$_2$ molecules adsorbed within
the IC's of nanotube bundles are shown. The inset shows the behavior
of $V_{\mathrm{eff}}^{(2)}$ close to the sign-reversal point.

Fig.$2$. Same as Fig.$1$ for He.

Fig.$3$. Same as Fig.$1$ for Ne. The inset shows that there is no reversal
in sign for $V_{\mathrm{eff}}^{(2)}$ in this case.

Fig.$4$.Angle-averaging the 27 orientations
of the equilateral triangle arrangement of (H$_2$)$_3$ to obtain
an isotropic potential, show in the bold curve with filled circles.
The thin curves represent each of the 10 distinct orientations 
at each separation.

Fig.$5$ .Hartree-Fock plus triple-dispersion fit of equilateral
and linear isotropic potentials.  Calculated data are represented
in symbols (filled = equilateral, empty = linear).  Solid curves
denote fits to equilateral geometry; dashed curves are fits to
linear geometry.

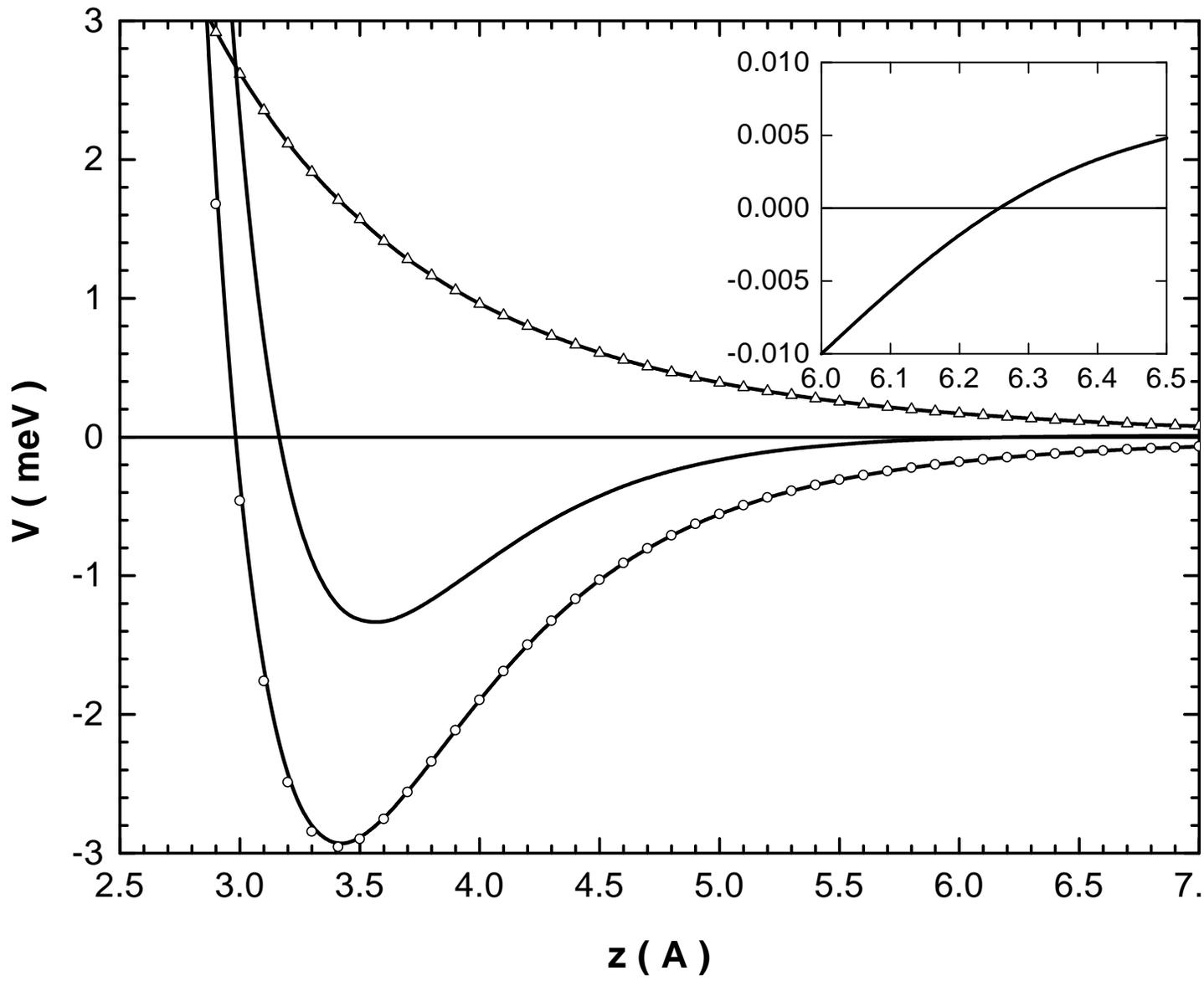
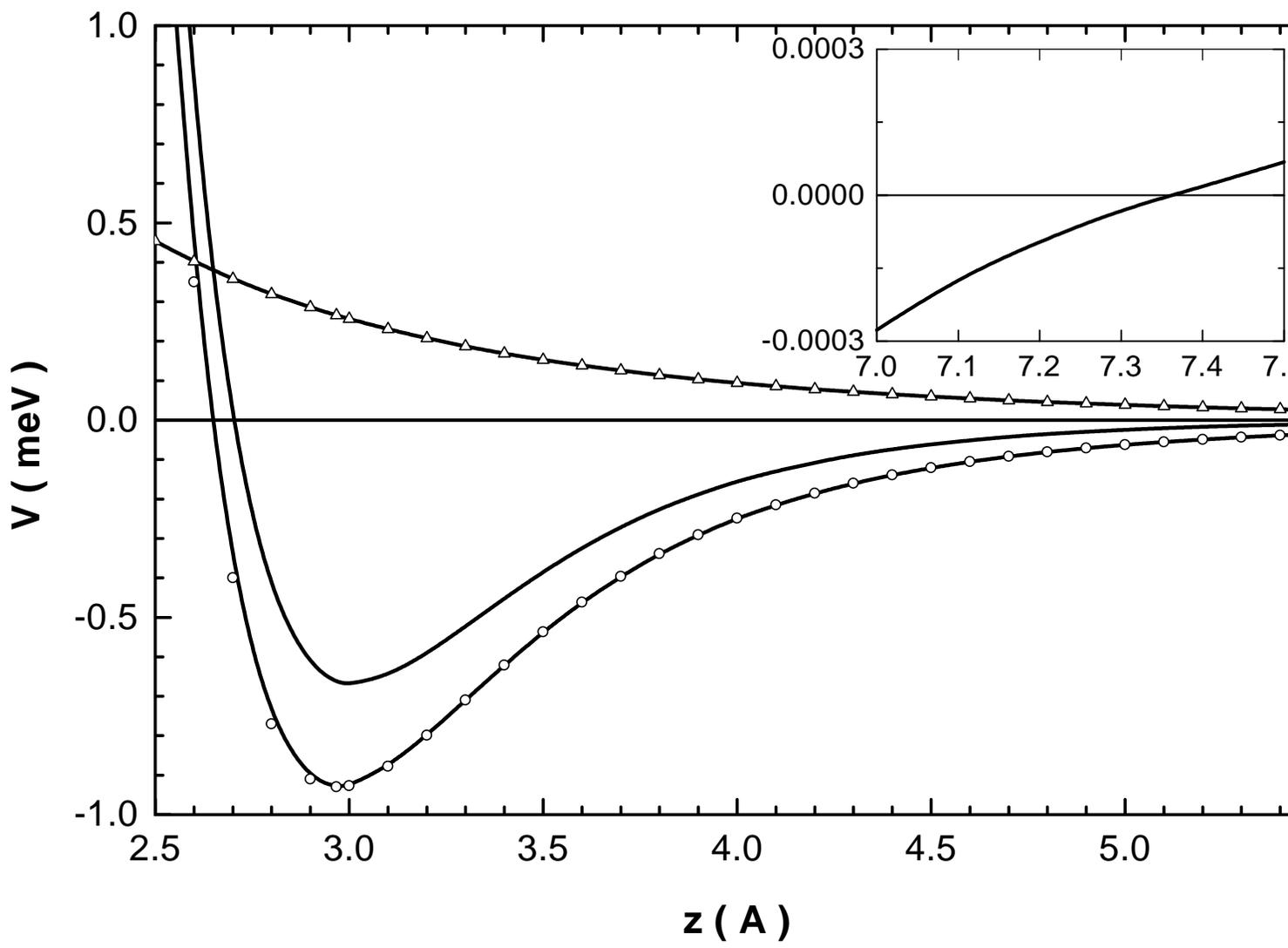
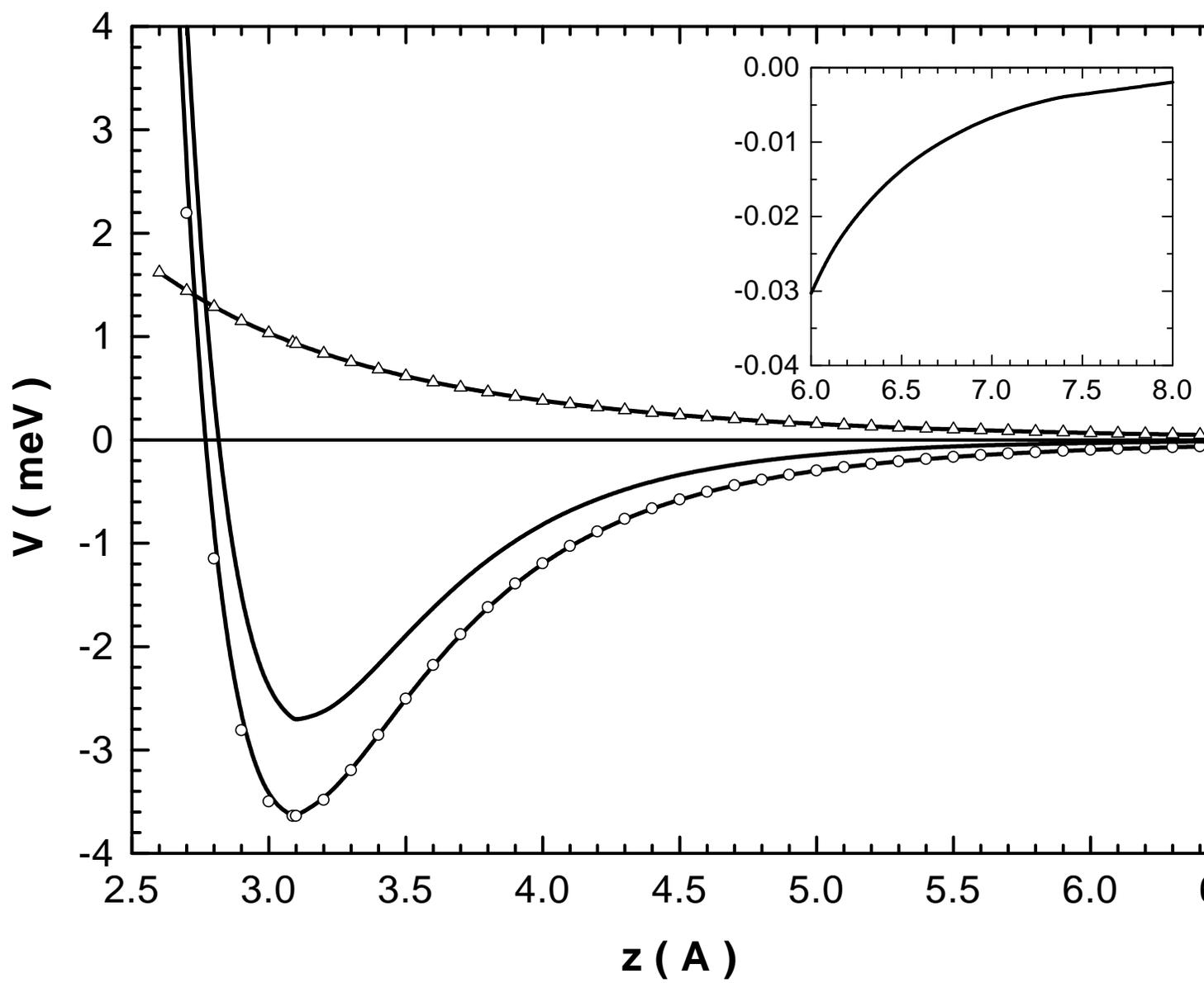
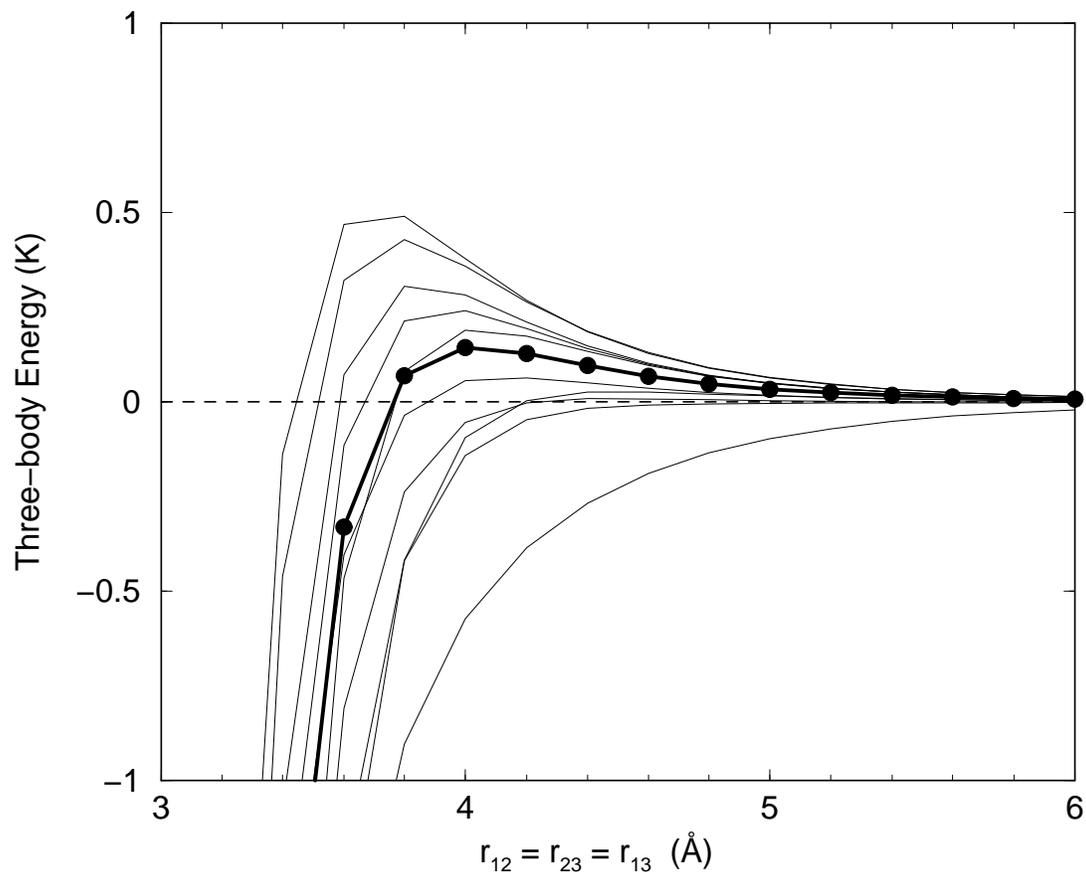
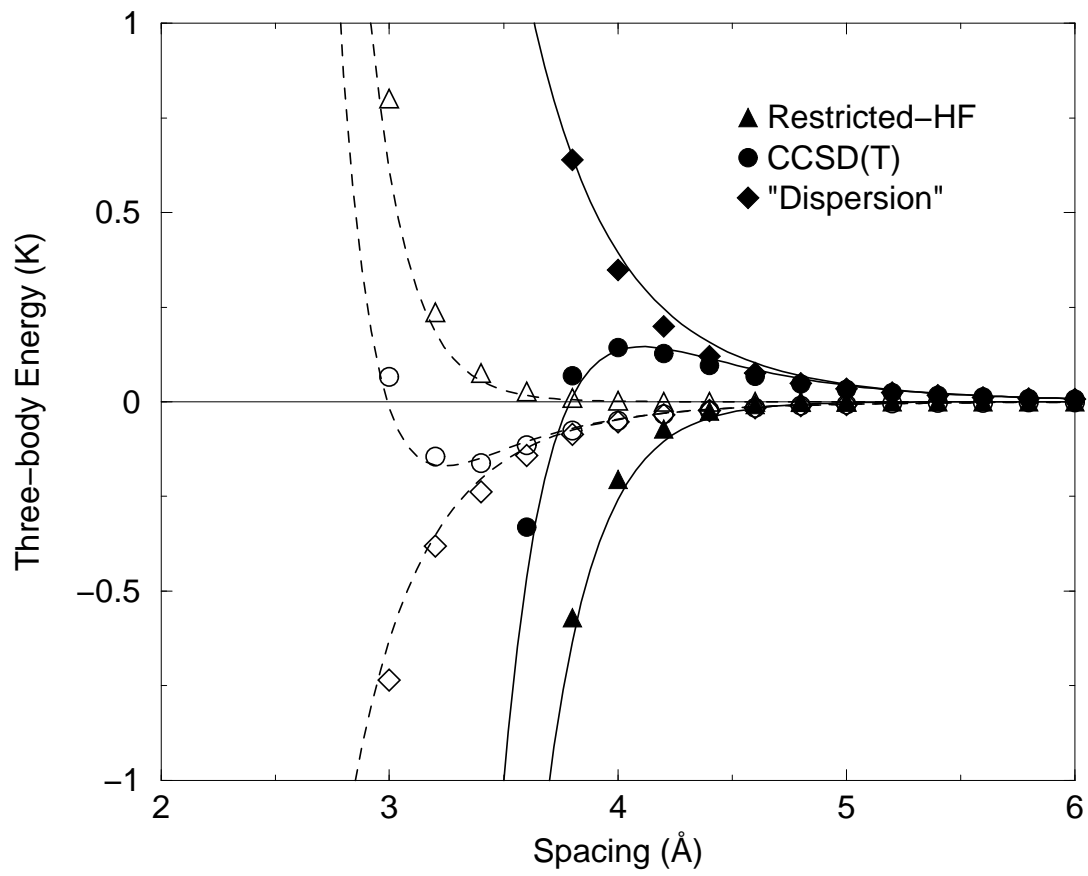
\end{document}